\documentclass[pra,aps,twocolumn]{revtex4-1}
\usepackage{amsmath}
\usepackage{amssymb}
\usepackage{graphicx}
\usepackage{bbold}
\usepackage{csquotes}

\newcommand{\beqa}{\begin{eqnarray}}
\newcommand{\eeqa}{\end{eqnarray}}

\begin{document}

\title{Time invariant $\mathcal{PT}$-product and phase locking in $\mathcal{PT}$-symmetric lattice models}
\author{Yogesh N. \surname Joglekar, Franck \surname Assogba Onanga, and Andrew K. \surname Harter}
\affiliation{Department of Physics, Indiana University Purdue University Indianapolis (IUPUI), Indianapolis, Indiana 46202 USA}
\date{\today}
\begin{abstract}
Over the past decade, non-Hermitian, $\mathcal{PT}$-symmetric Hamiltonians have been investigated as candidates for both, a fundamental, unitary, quantum theory, and open systems with a non-unitary time evolution. In this paper, we investigate the implications of the former approach in the context of the latter. Motivated by the invariance of the $\mathcal{PT}$ (inner) product under time evolution, we discuss the dynamics of wave-function phases in a wide range of $\mathcal{PT}$-symmetric lattice models. In particular, we numerically show that, starting with a random initial state, a universal, gain-site location dependent locking between wave function phases at adjacent sites occurs in the $\mathcal{PT}$-symmetry broken region. Our results pave the way towards understanding the physically observable implications of time-invariants in the non-unitary dynamics produced by $\mathcal{PT}$-symmetric Hamiltonians. 
\end{abstract}
\maketitle


\section{Introduction}
\label{sec:intro}
A standard axiom of quantum theory is that any observable is Hermitian operator~\cite{qmbasics}. Its Hermiticity ensures that its eigenvalues - experimentally observable quantities - are real and that the corresponding eigenvectors form a complete, orthonormal set. Among the observables, the Hamiltonian of a system is special, being the generator of its time evolution. When it is Hermitian, the Hamiltonian leads to a unitary time evolution. In 1998, Bender and Bottcher presented a broad class of non-Hermitian Hamiltonians for a non-relativistic particle on an infinite line with purely real spectra~\cite{bender1}. Although not Hermitian, they are invariant under combined operations of parity ($\mathcal{P}$) and time-reversal ($\mathcal{T}$). A typical parity-time ($\mathcal{PT}$) symmetric Hamiltonian consists of a Hermitian part and an anti-Hermitian part, i.e., $H_{PT}(\gamma)=H_0+i\gamma V$. The symmetry of the Hamiltonian $H_{PT}$ implies that the potential $V$ is an odd function. The spectrum of the Hamiltonian changes from purely real to complex-conjugate pairs when its non-Hermiticity $\gamma$ is increased. This transition is called $\mathcal{PT}$-symmetry breaking transition and occurs at the threshold $\gamma_{PT}$. Since the eigenfunctions of $H_{PT}(\gamma)$ are not orthogonal when $\gamma>0$, the time evolution generated by $H_{PT}$ is not unitary even if its spectrum is real~\cite{bender2,ajp,bender3}. Thus, when a system evolves under $H_{PT}$, the Dirac norm of a generic state fluctuates with time, and so do the inner-products between different states~\cite{review}. 

Following the discovery of $\mathcal{PT}$ symmetric Hamiltonians, the research has progressed along two non-overlapping lines. The first approach, intensely pursued initially, was to develop a complex extension of quantum mechanics~\cite{bender2}. Since $[\mathcal{PT},H_{PT}]=0$, the $\mathcal{PT}$-product, defined as $\langle\phi(t)|\psi(t)\rangle_\mathcal{PT}\equiv(\mathcal{PT}|\phi(t)\rangle)^T|\psi(t)\rangle$, remains constant with time~\cite{ajp}. However, it is not positive-definite and therefore cannot be used to construct a self-consistent quantum theory. This obstacle is circumvented by construction of a new, Hamiltonian-dependent, commuting operator $\mathcal{C}$ such that the $\mathcal{CPT}$-product, defined analogously, is positive-definite and gives rise to a unitary time evolution~\cite{ajp,bender3}. This approach, establishing the pseudo-Hermiticity~\cite{mostafa,mostafa1,mostafa2,mostafa3} or crypto-Hermiticity~\cite{mz1,mz2} of the non-Hermitian Hamiltonian, has been extensively investigated also in the language of general intertwining operators. It is valid only in the $\mathcal{PT}$-symmetric region, i.e., when the eigenvalues of $H_{PT}$ are purely real. The resultant unitary evolution allows one to study a host of traditional problems in the context of $\mathcal{PT}$-symmetric Hamiltonians. These include the quantum brachistocrone problem~\cite{qb1,qb2}, $\mathcal{PT}$-symmetric thermodynamics~\cite{ptthermo1}, Jarzynski inequality for $\mathcal{PT}$-symmetric Hamiltonians~\cite{ptthermo2}, and so on. However, the costs of redefining the inner product are that, in general, the expectation values of "usual" observables such as position $\hat{x}$, momentum $\hat{p}$, or spin projections $(\sigma_x,\sigma_y,\sigma_z)$ are not real, and such unitary time evolutions violate no-signaling principle~\cite{causality}. Therefore, predictions obtained from a unitary evolution via the $\mathcal{CPT}$ inner-product do not correspond to experimental results, although they can be simulated via experimental, circuit-model quantum simulators~\cite{ptqs}. 

The second approach started about a decade ago~\cite{th1,th2}. It treats $H_{PT}$ as an {\it effective} Hamiltonian for an open system - classical or quantum - where a unitary time evolution is neither required not expected. In this interpretation, the antisymmetric potential $i\gamma V$ represents a gain for the system when $V>0$ and is accompanied by an equal loss with $V<0$ at the parity symmetric location. The development of complex-conjugate eigenvalues for $H_{PT}$ then denotes the emergence of two eigenmodes, one of which amplifies with time and the other one decays. This approach does not change the fundamental, Dirac inner product of quantum theory, and is applicable in both $\mathcal{PT}$-symmetric and $\mathcal{PT}$-broken (complex conjugate spectrum) phases. It has been immensely successful in predicting a multitude of novel phenomena in $\mathcal{PT}$-symmetric systems and explaining the subsequent experimental observations~\cite{expt1,expt2,expt3,uni1,uni2,expt4,expt5,expt6,expt7,expt8,expt9}. However, in this case, due to the non-unitary time evolution generated by the effective Hamiltonian, the constants of motion for such a system~\cite{ptdimerib} are not clear.  

In this paper, we bridge the gap between these two approaches by connecting the time invariant $\mathcal{PT}$-product mentioned in the first approach with the non-unitary evolution of a physical system in the second approach. We show that, deep in the $\mathcal{PT}$-broken state, the exponentially-in-time growth of the Dirac norm, combined with the constant-in-time constraint on the $\mathcal{PT}$-product (and expectation values of other intertwining operators), {\it leads to phase locking that is independent of the initial state.} 

The plan of the paper is as follows. In Sec.~\ref{sec:twothree}, we introduce the general notation, obtain analytical results for a $\mathcal{PT}$-symmetric dimer and trimer, and thus elucidate our motivation for focusing on the temporal dynamics of adjacent-site phase differences of a wave function. Section~\ref{sec:obc} has numerical results for the time-dependent adjacent-site phase differences obtained for a wide variety of $N$-site one-dimensional, tight-binding chains with one or more pairs of $\mathcal{PT}$-symmetric gain and loss potentials. Corresponding results for lattices with periodic boundary conditions are presented in Sec.~\ref{sec:pbc}. We conclude the paper with a brief discussion in Sec.~\ref{sec:disc}. 


\section{PT symmetric dimer and trimer}
\label{sec:twothree} 

Let us consider a $\mathcal{PT}$-symmetric dimer, represented by two sites $|1\rangle$ and $|2\rangle$ having tunneling amplitude $J>0$ and gain-loss potentials $\pm i\gamma$. Its Hamiltonian is given by $H_2=-J\sigma_x+i\gamma\sigma_z\neq H^\dagger_2$ where $\sigma_x,\sigma_z$ are standard Pauli matrices. Although not Hermitian, $H_2$ is invariant under combined parity $\mathcal{P}=\sigma_x$ and time-reversal $\mathcal{T}=*$ operations. The spectrum of $H_2$ undergoes  a $\mathcal{PT}$ transition at $\gamma_{PT}=J$ where the two eigenvalues $\pm\lambda_2=\pm\sqrt{J^2-\gamma^2}$ become degenerate as do the corresponding eigenfunctions. Since $H_2^2=\lambda_2^2$, the non-unitary time evolution operator $G_2(t)=\exp(-iH_2t)$ is given by 
\begin{equation}
\label{eq:gdimer}
G_2(t)=\cos(\lambda_2 t){\bf 1}_{2}-i(H_2/\lambda_2)\sin(\lambda_2 t).
\end{equation}
In the $\mathcal{PT}$-symmetric state, the norm of $G(t)$ oscillates in time. At the exceptional point $\gamma=J$, it grows linearly with time, and in the $\mathcal{PT}$-broken state it grows as $\exp(+\Lambda_2 t)$ where $\Lambda_2=\sqrt{\gamma^2-J^2}>0$. We start with a random initial state $|\psi(0)\rangle=\sum_k a_k e^{i\phi_k}|k\rangle$, where $|k\rangle$ represents the a state localized at site $k$, $a_k\geq 0$, $\phi_k\in(0,2\pi]$. (Note that for the purposes of our argument, the initial value of the Dirac norm of $|\psi(0)\rangle$ is irrelevant.) The time-invariance of the $\mathcal{PT}$ product $N_{PT}=\langle\psi(t)|\psi(t)\rangle_{PT}$ implies that 
\begin{equation}
\label{eq:npt2}
N_{PT}(t)=2a_1(t)a_2(t)\cos\left[\phi_{2}(t)-\phi_{1}(t)\right]=\mathrm{constant}.
\end{equation}
When $N_{PT}=0$, i.e., the initial state is localized only on one site or has an initial $\pi/2$ phase difference between the two sites, it follows that the phase difference between the two sites remains fixed, $\theta_{2}(t)\equiv\left[\phi_2(t)-\phi_1(t)\right]=\pm\pi/2$. This result is true in the Hermitian limit, $\gamma=0$, as well.  

When $N_{PT}\neq 0$, i.e., the initial state is distributed over the two sites and complex, then the time invariance of $N_{PT}$ implies that the phase difference $\theta_{2}(t)$ evolves with time and is not stationary. However, in the $\mathcal{PT}$-broken state, at long times $\Lambda_2t\gg1$, the on-site weights $a_k(t)$ grow exponentially and therefore time-invariance of Eq.(\ref{eq:npt2}) requires that $\cos\theta_{2}(t)\propto \cos\theta_2(0)\exp(-2\Lambda_2 t)\rightarrow 0^{\pm}$ depending on the initial sign of $N_{PT}$.  Thus the phase difference $\theta_{2}\rightarrow \pm\pi/2$ {\it for an arbitrary initial state.} This is in sharp contrast with the corresponding results in the Hermitian case or the $\mathcal{PT}$ symmetric region. A similar analysis follows for a $\mathcal{PT}$-symmetric trimer. The Hamiltonian for the trimer is given by $H_3=-JS_x+i\gamma S_z$ where $S_x$ and $S_z$ are spin-1 representations of angular momentum operators~\cite{ptrimer}. Its eigenvalues are given by  $\{0,\pm\lambda_3\}$ with $\lambda_3=\sqrt{J^2-\gamma^2}$; they become degenerate at $\gamma_{PT}=J$ and complex for a larger $\gamma$. The invariance of $\mathcal{PT}$ product in this case implies that 
\begin{equation}
\label{eq:npt3}
a_2^2(t)+2a_1(t)a_3(t)\cos\left[\phi_3(t)-\phi_1(t)\right]=\mathrm{constant}.
\end{equation}
Eq.(\ref{eq:npt3}) contains information about the wave function phases only at the gain and loss sites, but not about the central, neutral site. Thus, we cannot obtain any definitive conclusions about the phase-locking in the $\mathcal{PT}$-broken state based solely on Eq.(\ref{eq:npt3}). Motivated by the phase-locking result in Eq.(\ref{eq:npt2}), in the following sections, we numerically investigate the fate of adjacent-site phase differences $\theta_k(t)\equiv\left[\phi_k(t)-\phi_{k-1}(t)\right]$ for random initial wave functions, when the system is in the $\mathcal{PT}$-broken state.  


\section{Lattices with open boundary conditions}
\label{sec:obc}

In this section, we consider the time evolution of $\theta_k(t)$ in a wide variety of one-dimensional, $N$ site, tight-binding lattices with open boundary conditions. 


\subsection{Uniform tunneling chain}
\label{subsec:chain}

\begin{figure*}[t]
\includegraphics[width=\columnwidth]{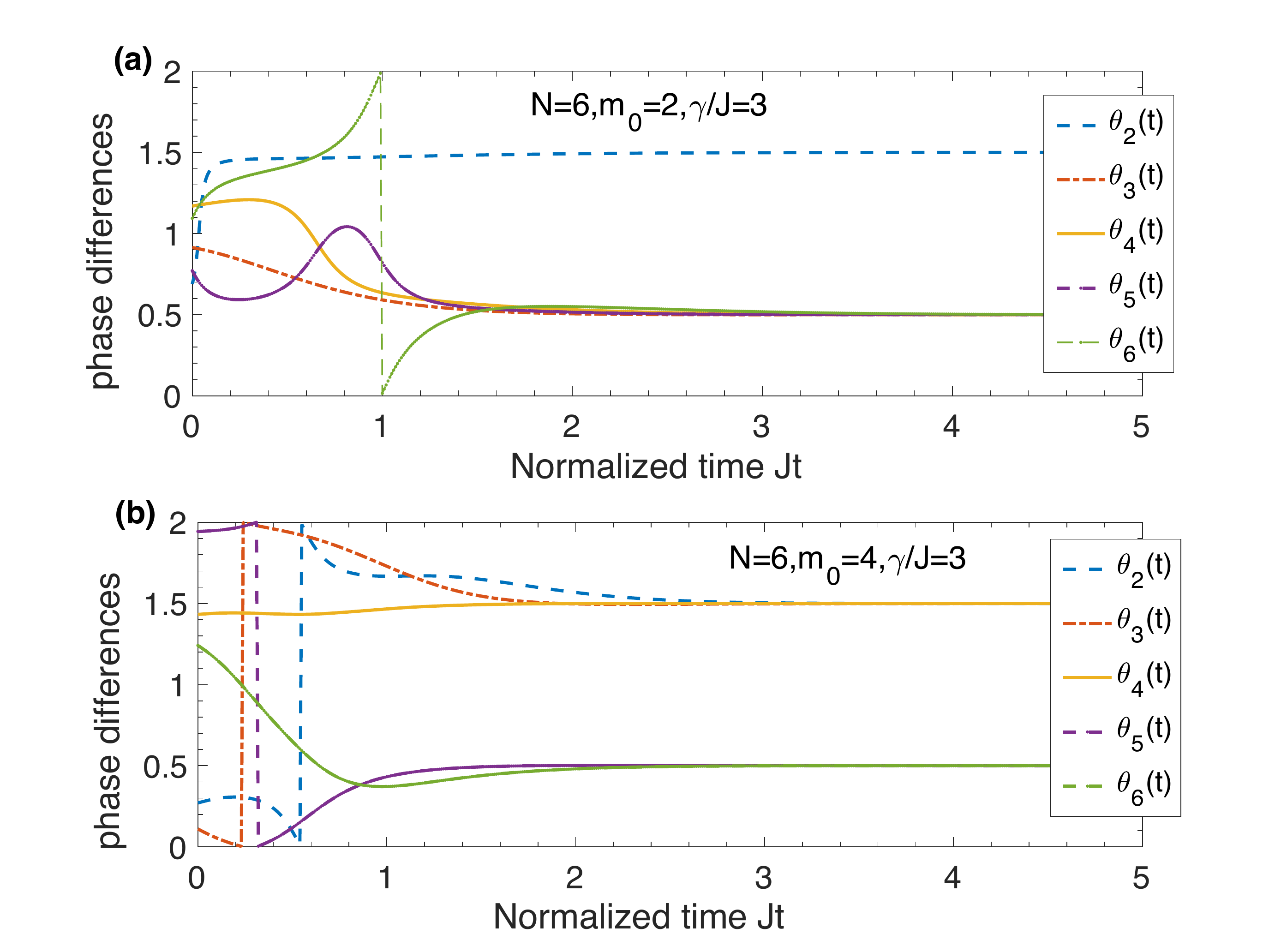}
\includegraphics[width=\columnwidth]{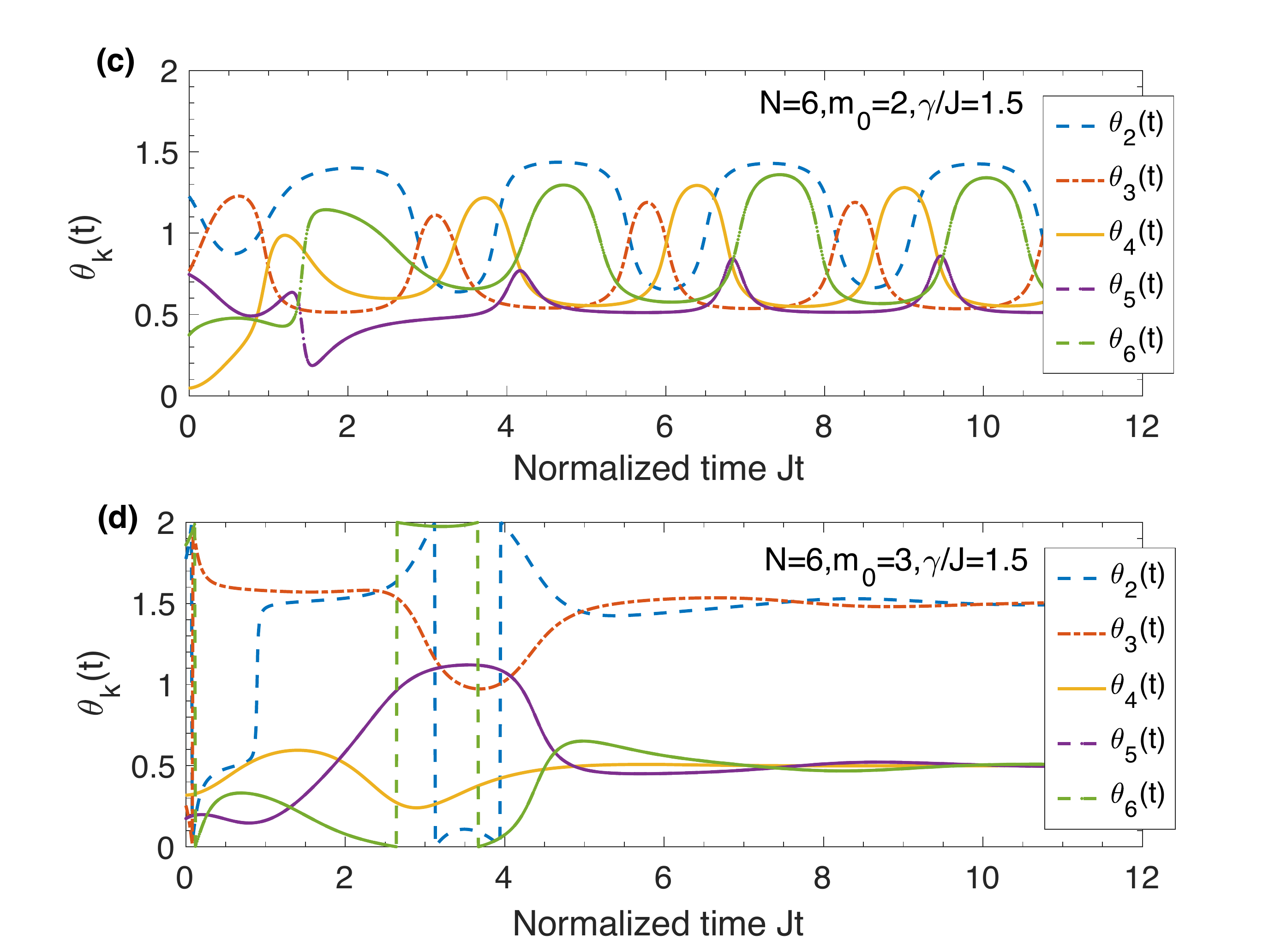}
\caption{Time evolution of the adjacent site phase differences $\theta_k(t)$, measured in units of $\pi$,  for an $N=6$ site lattice. (a), (b) deep in the $\mathcal{PT}$-broken region, $\gamma/J=3$, the phases $\theta_{k\leq m_0}$ saturate to $3\pi/2$ and $\theta_{k>m_0}$ saturate to $\pi/2$. (c) the phase differences $\theta_k(t)$ reach steady state value if and only if $\gamma\geq\gamma_{PT}$ is large enough so that all eigenvalues that are supposed to become complex have done so. (d) changing the gain location to $m_0=N/2=3$ maximally breaks the $\mathcal{PT}$-symmetry and thus leads to steady-state. These results are independent of the random initial states used, as well as the exact value of the gain-loss strength as long as the criteria mentioned above are satisfied.}
\label{fig:chain}
\end{figure*}

Let us consider an $N$-site chain with constant nearest neighbor tunneling amplitude $J>0$ and a pair of gain and loss potentials $\pm i\gamma$ located at mirror symmetric sites $m_0$ and $\bar{m}_0=N+1-m_0$. The $\mathcal{PT}$-symmetric, non-Hermitian Hamiltonian for the chain with open boundary conditions is given by 
\begin{eqnarray}
\label{eq:hpt}
H_{c}(\gamma) &= &-J\sum_{k=1}^{N-1}\left( |k\rangle\langle k+1| + |k+1\rangle\langle k|\right) \nonumber\\
& + & i\gamma\left( |m_0\rangle\langle m_0| - |\bar{m}_0\rangle\langle\bar{m}_0|\right),
\end{eqnarray}
where $|k\rangle$ denotes a single-particle state localized at site $k$. The parity operator on such a lattice is given by $\mathcal{P}:k\rightarrow \bar{k}$, and the time-reversal operator corresponds to complex conjugation, $\mathcal{T}=*$. The $\mathcal{PT}$-breaking threshold $\gamma_{PT}(m_0)$ shows a U-shape profile. It is maximum at $\gamma_{PT}\sim J$ when the gain and loss are farthest apart~\cite{song}, is algebraically suppressed when $m_0/N\sim 1/4$, and is enhanced again to $\gamma_{PT}\sim J$ when the gain and loss are nearest neighbors~\cite{mark} Starting from an initial random state $|\psi(0)\rangle$, the time-evolved state is given by $|\psi(t)\rangle=G_c(t)|\psi(0)\rangle=\sum_{k=1}^N a_k(t) e^{i\phi_k(t)}|k\rangle$. Here $G_c(t)=\exp(-iH_{c}t)$ is the non-unitary time evolution operator. We keep in mind a waveguide (or a resonator) array with amplification in waveguide $m_0$ and an equal loss in its mirror-symmetric waveguide $\bar{m}_0$ as a possible realization of this chain. Thus $a_k(t)\exp\left[i\phi_k(t)\right]$ denotes the time-dependent complex amplitude of the slowly varying envelope of the electric field in waveguide $k$. Since $H_{c}\neq H_{c}^\dagger$, it follows that the Dirac norm of the wave function $\langle\psi(t)|\psi(t)\rangle=\sum_{k=1}^N a_k^2(t)$ is not a time invariant. In fact it  undergoes bounded oscillations in the $\mathcal{PT}$-symmetric state, has a power-law growth at the $\mathcal{PT}$-breaking transition point, and has an exponentially increasing envelope in the $\mathcal{PT}$-broken state. In contrast, the time-invariant $\mathcal{PT}$ product of the state $|\psi(t)\rangle$ with itself is given by 
\begin{equation}
\label{eq:ptproduct}
N_{c}=\langle\psi(t)|\psi(t)\rangle_{\mathcal{PT}}=\sum_{k=1}^N a_k(t) a_{\bar{k}}(t) e^{i\left[\phi_k(t)-\phi_{\bar{k}}(t)\right]}.
\end{equation}

Figure~\ref{fig:chain} shows the numerically obtained time evolution of the $N-1$ adjacent-site phase differences $\theta_{k}(t)$ for an $N=6$ site chain with different locations $m_0$ and strengths of the gain potential $+i\gamma$. Figure~\ref{fig:chain}a shows that for $m_0=2$ and $\gamma/J=3\gg \gamma_{PT}(m_0)$~\cite{mark}, only $\theta_2(t)\rightarrow 3\pi/2$ whereas for all other sites $k$ beyond the gain site $m_0=2$, $\theta_k(t)\rightarrow\pi/2$. Figure~\ref{fig:chain}b shows the results for the gain location at $\bar{m}_0=4$. We remind the reader that in this case, the $\mathcal{PT}$-breaking threshold is the same. We see, however, that now $\theta_2(t),\theta_3(t)$ and $\theta_4(t)$ all approach the value $3\pi/2$, and $\theta_k(t)\rightarrow\pi/2$ only for $k>m_0=4$. These results are independent of the random initial state, as well as the exact value of the gain-loss strength $\gamma$. 

Figure~\ref{fig:chain}c shows the temporal dynamics for an $N=6$ lattice with gain at $m_0=2$ and a gain-loss strength of $\gamma/J=1.5$. The phase differences $\theta_k(t)$, after an initial transient, show periodic oscillations that do not reach a steady state value. This behavior, showing both transient and periodic features, is only found when the spectrum is not purely real, $\gamma>\gamma_{PT}$, and yet, $\gamma$ is not large enough so that all eigenvalues that are supposed to become degenerate and complex have become so~\cite{mark}. Figure~\ref{fig:chain}d shows corresponding results for the same lattice, with the same gain-loss strength, but with nearest-neighbor gain and loss potentials, $m_0=N/2=3$. We see that $\theta_2(t),\theta_3(t)\rightarrow3\pi/2$ whereas all other phases $\theta_k(t)\rightarrow\pi/2$ for $k>m_0$. We remind the reader that when $m_0=N/2$, the $\mathcal{PT}$-symmetry is maximally broken and all eigenvalues becomes complex simultaneously at $\gamma_{PT}=J$~\cite{jake}. 

Results in Fig.~\ref{fig:chain} show that deep in the $\mathcal{PT}$-broken state, {\it the chain becomes phase-locked, with the gain-site marking the location where phase-locked value changes from $3\pi/2$ to $\pi/2$.}


\subsection{Periodic tunneling (SSH and AAH) chain}
\label{subsec:sshaah}

\begin{figure}
\includegraphics[width=\columnwidth]{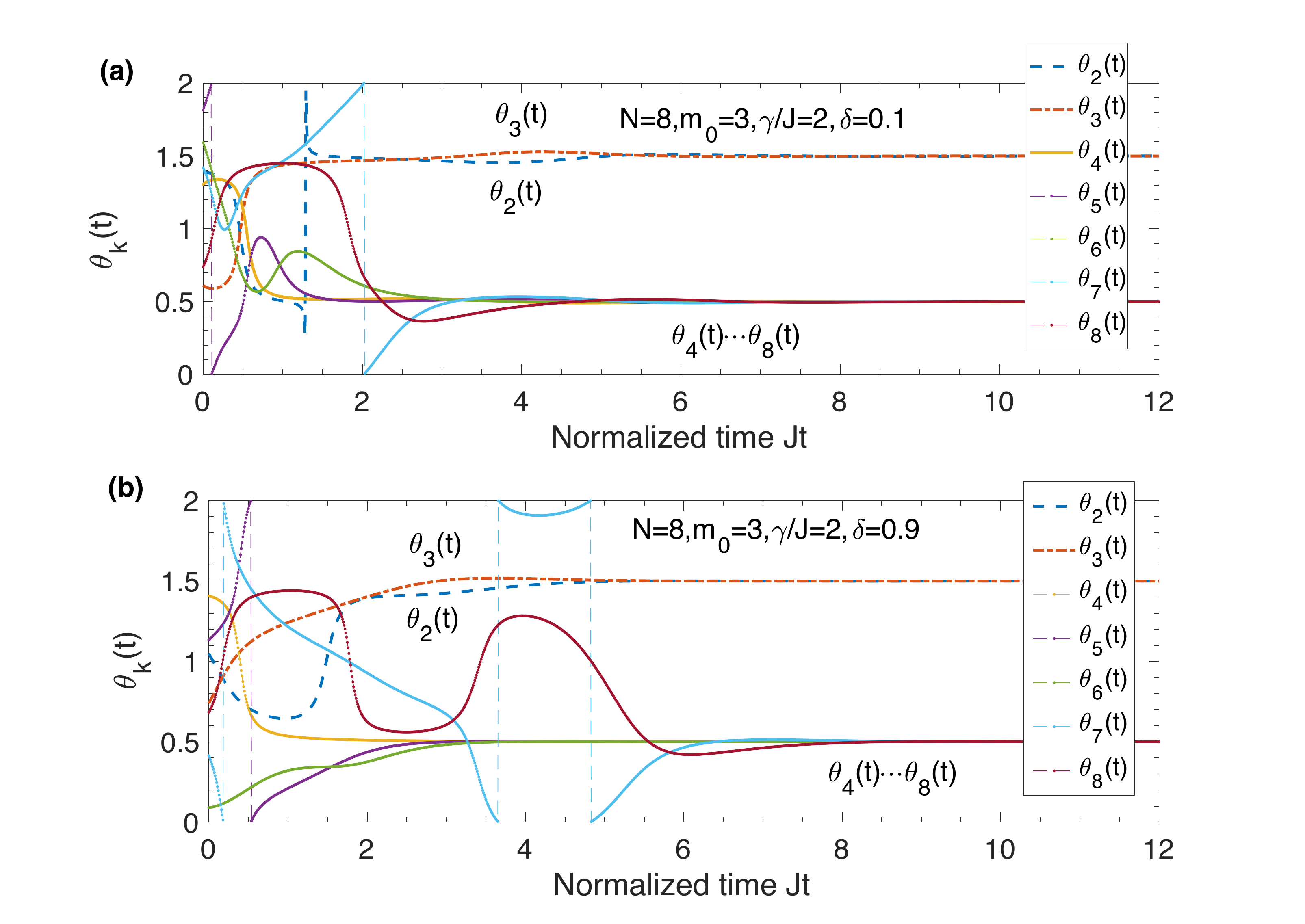}
\caption{Adjacent-site phase-difference dynamics in an $N=8$ SSH chain with gain potential $\gamma=2J$ at site $m_0=3$ shows the same phase locking phenomenon seen in Fig.~\ref{fig:chain}. 
(a) for a small tunneling differential, $\delta=0.1$, the phase differences $\theta_2(t)$ and $\theta_3(t)$ saturate to $3\pi/2$ whereas the rest, $\theta_4(t),\cdots,\theta_8(t)$ saturate to $\pi/2$. (b) a larger tunneling differential, $\delta=0.9$, leads to a slower approach to saturation.}
\label{fig:ssh}
\end{figure}
Next, we consider an open $N$-site chain with a site-dependent tunneling amplitude $J(k)$ in the presence of a single pair of gain-loss potentials $\pm i\gamma$ at locations $m_0,\bar{m}_0$ respectively. The Su-Schrieffer-Heeger (SSH) model is given by a period-2 tunneling profile,
\begin{equation}
\label{eq:ssh}
J(k)=\left\{\begin{array}{ccc}
J & & k=0\mod 2,\\
J(1-\delta) & & k=1\mod 2,
\end{array}\right.
\end{equation}
where $\delta\leq 1$ quantifies the strong bond vs. the weak bond~\cite{ssh1,ssh2}. When $N$ is even, this tunneling profile is parity symmetric and the $\mathcal{PT}$-symmetry breaking threshold $\gamma_{PT}(m_0)$ shows a U-shaped profile that depends, in detail, on the ratio of the two tunneling amplitudes~\cite{andrewssh}. Figure~\ref{fig:ssh} shows the time-evolution of the phases $\theta_k(t)$ for an $N=8$ site lattice with gain location $m_0=3$ and gain-loss strength $\gamma/J=2$, with tunneling differentials $\delta=0.1$ (a) and $\delta=0.9$ (b) respectively.  In both cases, the adjacent-site phase difference saturates, i.e., $\theta_{k\leq m_0}(t)\rightarrow3\pi/2$ and $\theta_{k>m_0}(t)\rightarrow\pi/2$. However, the critical value of $\gamma$ at which the spectrum becomes maximally complex increases as the tunneling differential $\delta$ increases. Therefore, at a constant loss-strength $\gamma/J=2$, we see a slower approach to saturation in Fig.~\ref{fig:ssh}b relative to that in Fig.~\ref{fig:ssh}a. The results obtained here are independent of the random initial state chosen, the gain-loss strength, and the value of $\delta$. These variables - primarily the latter two variables - only determine the time needed to reach the saturation value. 

\begin{figure}
\includegraphics[width=\columnwidth]{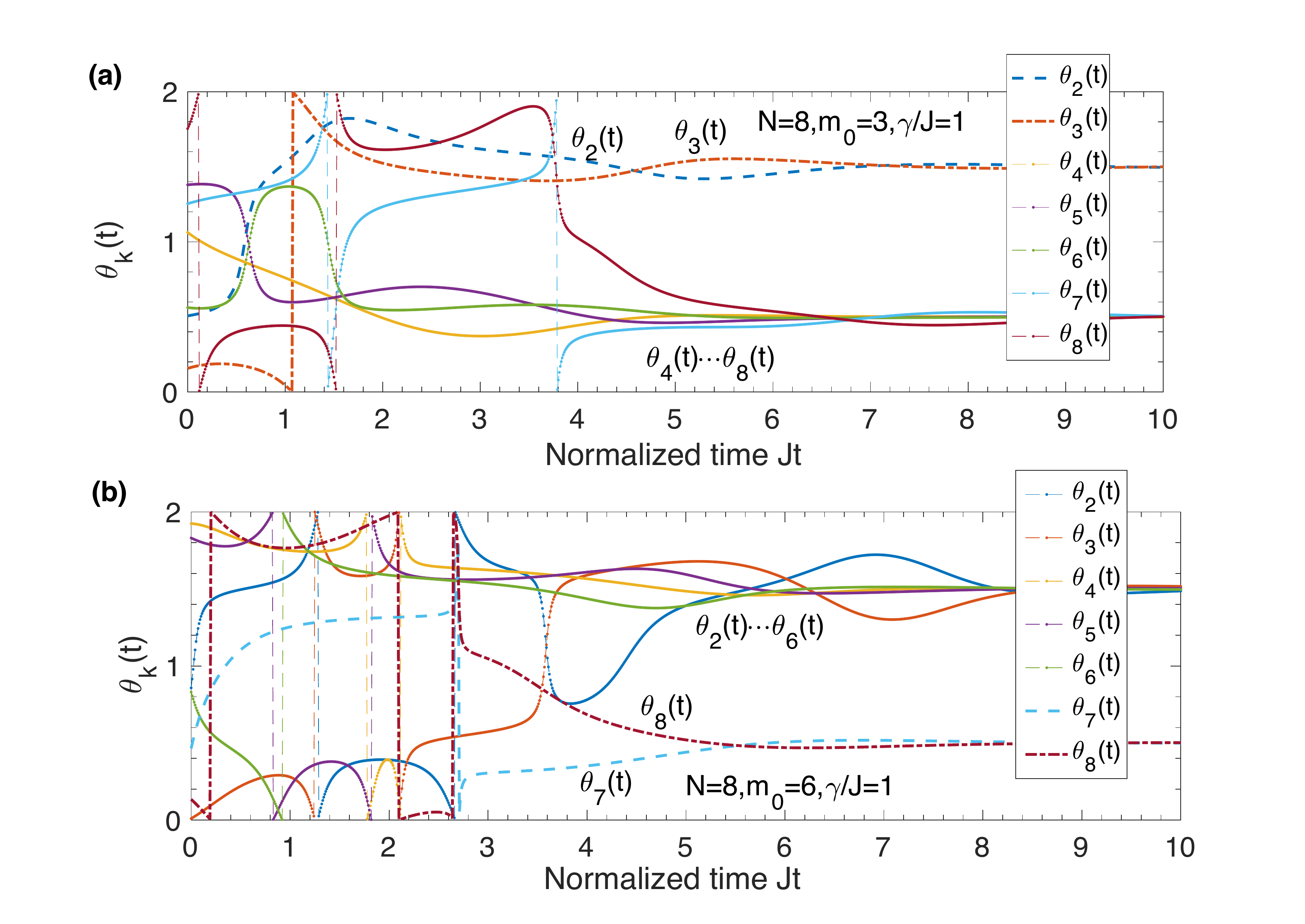}
\caption{Phase difference dynamics in an AAH chain with $N=8$ and period 3 tunneling profile, $\{J_1,J_2,J_3\}=J\{1,0.8,0.4\}$. The gain-loss strength is $\gamma/J=1$. (a) when $m_0=3$, phases $\theta_4(t),\cdots,\theta_8(t)$ saturate to $\pi/2$. (b) when $m_0=6$, the parity symmetric location, the phases $\theta_2(t),\cdots,\theta_6(t)$ saturate to $3\pi/2$. Although the AAH chain is not $\mathcal{PT}$ symmetric, it has a finite threshold~\cite{aah,sr}.}
\label{fig:aah}
\end{figure}
When the tunneling profile has a periodicity $p\geq3$, we get the Aubrey-Andre Harper (AAH) model~\cite{aah1,aah2}. For example, when $p=3$, the tunneling profile $J(k)$ repeats as $J_1,J_2,J_3,J_1,\cdots$. For $p\geq 3$ {\it this tunneling profile is not parity symmetric for any lattice size.} Nonetheless, due to the hidden symmetry of eigenfunctions of the AAH tunneling Hamiltonian, such a lattice has a finite $\mathcal{PT}$-symmetry breaking threshold when $N+1=0\mod p$ and $m_0=0\mod p$~\cite{aah,sr}. Figure~\ref{fig:aah} shows the time-evolution of $\theta_k(t)$ for an $N=8$ AAH model with period $p=3$. When the gain location is $m_0=3$ and the loss site is $\bar{m}_0=6$, the $\theta_2(t),\theta_3(t)$ phases saturate to $3\pi/2$ whereas the rest saturate to $\pi/2$ (panel a). On the other hand, when the gain site is at $m_0=6$, the phase differences $\theta_2(t),\ldots,\theta_6(t)$ all approach $3\pi/2$ and the remaining two saturate to a value of $\pi/2$ (panel b). 

These results show that deep in the $\mathcal{PT}$-broken state, generically, the wave function phases between adjacent sites are locked at $3\pi/2$ up to the location of the gain site, and then they switch to being locked at $\pi/2$. We emphasize that this phase-locking is different from the $\pi/2$-phase shift for wave function amplitudes in a Hermitian, tight-binding array~\cite{silberberg}. The latter arise solely when the initial state is confined to a single site. Our result is applicable to arbitrary initial states. 


\subsection{Perfect state transfter (PST) chain}
\label{subsec:pst}

\begin{figure}
\includegraphics[width=\columnwidth]{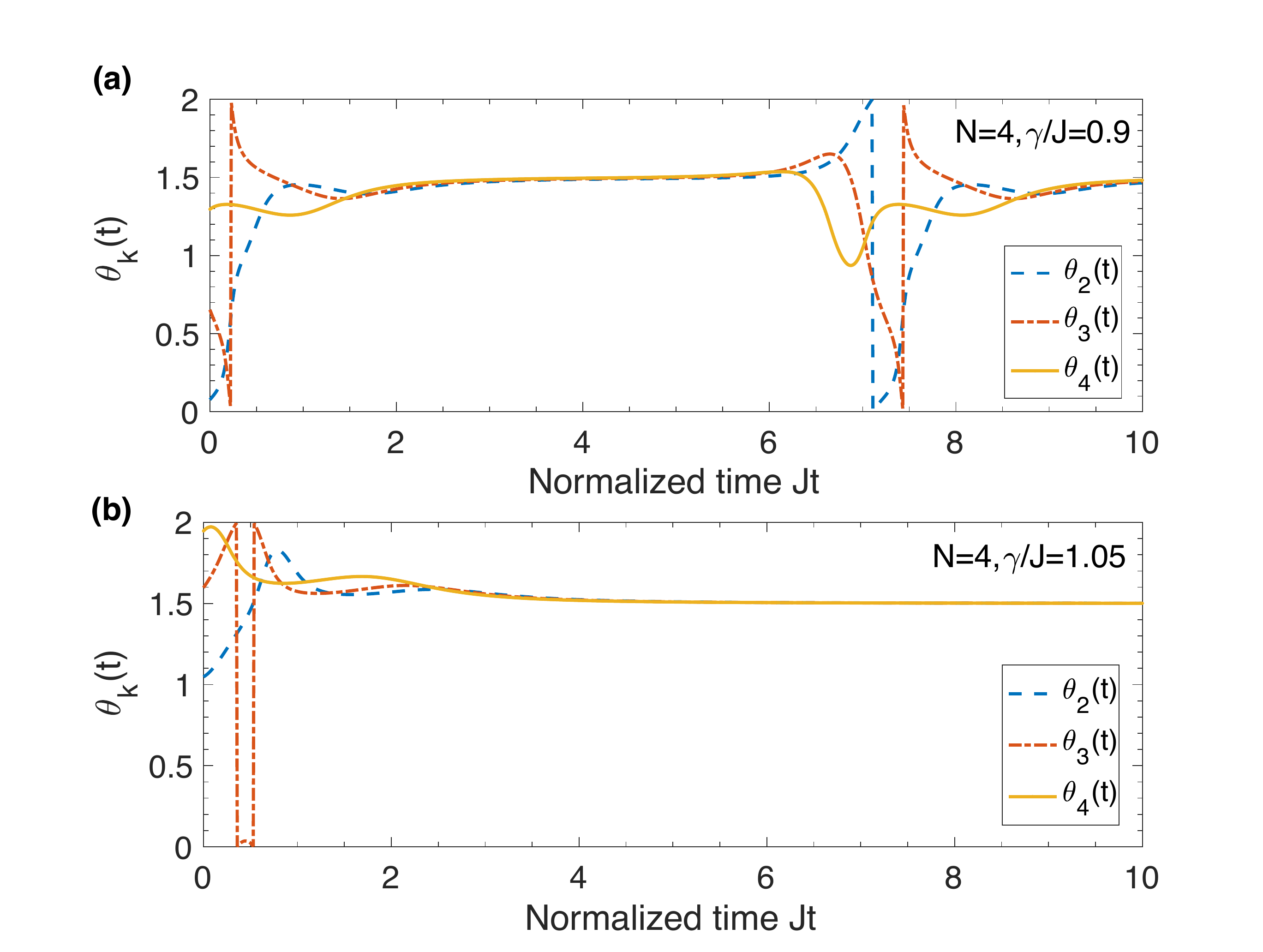}
\caption{Phase-difference dynamics in an $N=4$ perfect-state transfer lattice. (a) when gain-loss strength is smaller than the threshold, $\theta_k(t)$ oscillate with period $T=\pi/\sqrt{J^2-\gamma^2}$. (b) for $\gamma>\gamma_{PT}$, all the phases saturate to $3\pi/2$. These results are generic, and show that $\theta_{k}\rightarrow3\pi/2$ for any PST lattice with gain region in the second-half of the lattice. For gain region in the first half of the lattice, $\gamma<0$, all phases saturate to $\pi/2$.}
\label{fig:pst}
\end{figure}
In the past subsections, we only considered finite chains with a single pair of gain and loss potentials. In this subsection, we generalize our results to a chain with extended $\mathcal{PT}$-symmetric gain loss potentials. In general, an extended gain-loss potential profile leads to an algebraically fragile $\mathcal{PT}$-symmetric phase~\cite{mark,bagchi,divergent}, and therefore, such models are not particularly interesting. However, a $\mathcal{PT}$ symmetric perfect-state-transfer (PST) model is a notable exception. Let us consider the Hamiltonian $H_{pst}(\gamma)=-JS_x+i\gamma S_z$ where $S_x,S_z$ are now spin $S=(N-1)/2$ dimensional representations of the angular momentum algebra, and the site index $k$ in this $N=2S+1$-site chain maps on to the $S_z$ angular momentum projection index. 

The Hamiltonian $H_{pst}$ represents a chain where the parity-symmetric tunneling between sites $k$ and $k+1$ is equal to $\sqrt{k(N-k)}/2$~\cite{sugato,clint}, and the $\mathcal{PT}$-symmetric gain-loss potential linearly changes from $-i\gamma S$ to  $+i\gamma S$ in steps of $i\gamma$. The particle-hole symmetric, equidistant spectrum of $H_{pst}$ is given by $\epsilon_m= m\sqrt{J^2-\gamma^2}$ for $-S\leq m\leq S$, and it has a single exceptional point of order $N$ at the threshold $\gamma_{PT}=J$. 

Figure~\ref{fig:pst} shows the fate of the adjacent-site phase differences $\theta_k(t)$ for an $N=4$ PST chain. Panel a shows that when the gain-loss strength is below the threshold, $\gamma/J=0.9$, starting from a random initial state, the three phase differences $\theta_2(t),\theta_3(t),\theta_4(t)$ undergo periodic oscillations with period $T=\pi/\sqrt{J^2-\gamma^2}$. Panel b shows that for $\gamma/J=1.005$, when all eigenvalues have become complex-conjugate pairs, the phase differences reach steady state value, $\theta_k(t)\rightarrow 3\pi/2$. This is consistent with earlier observation that the phase differences saturate to $3\pi/2$ for all sites preceding and including the site that has the maximum gain potential. The results presented in Fig.~\ref{fig:pst} are true for any chain size $N$ and random initial states. We also find that all adjacent-site phase differences $\theta_k(t)$ saturate to $\pi/2$ when the sign of $\gamma$ is reversed, meaning the largest gain potential is on the first site. 

The results in this section show that the phase locking phenomenon presented here is robust in a wide variety of finite lattices with open boundary conditions. In the following section, we consider the effects of periodic boundary condition. 

\section{Lattices with periodic boundary conditions}
\label{sec:pbc} 

\begin{figure*}
\includegraphics[width=\columnwidth]{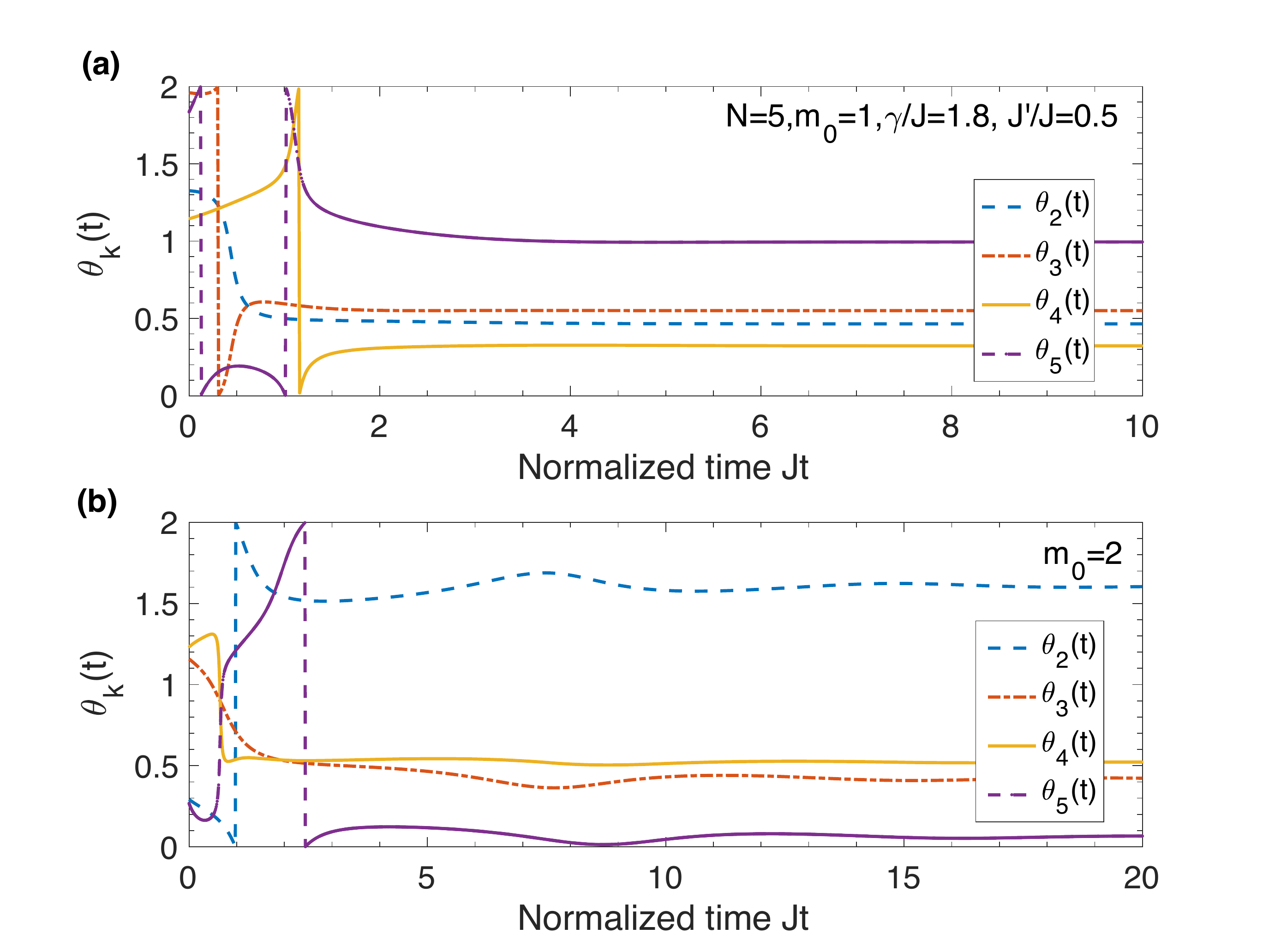}
\includegraphics[width=\columnwidth]{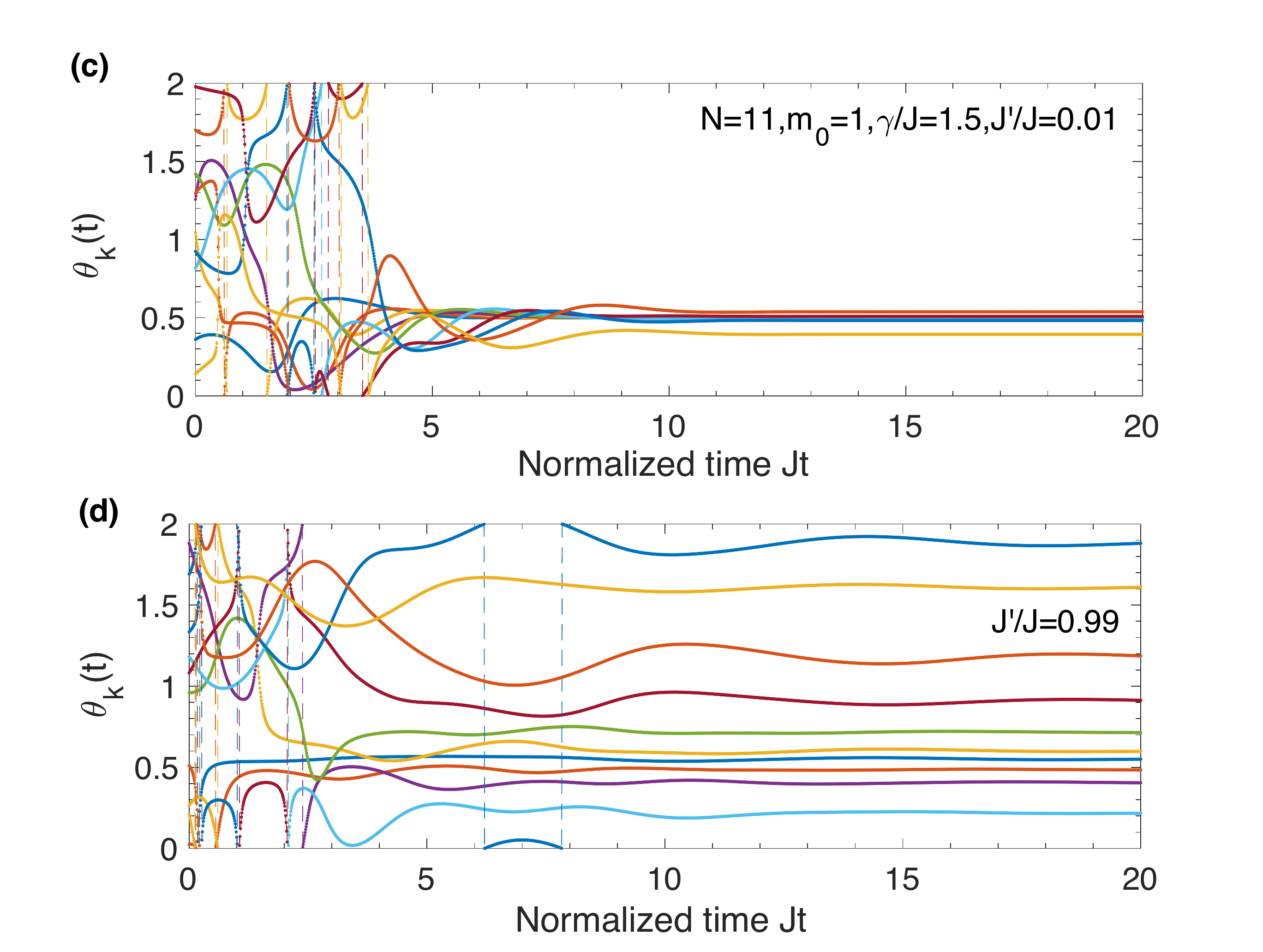}
\caption{Time evolution of $\theta_k(t)$ in a $\mathcal{PT}$-symmetric, two-tunneling ring. Results for an $N=5$ ring with $J'=0.5J$ and $\gamma/J=1.8$ show that the $\theta_k(t)$ saturate to multiple, different values that depend on the gain-site locations $m_0=1$ (a) and $m_0=2$ (b). Results for an $N=11$ ring with gain strength $\gamma/J=1.5$ at location $m_0=1$ shows that (c) when $J'\ll J$, the phase differences saturate to values near $\pi/2$ whereas (b) for $J'\sim J$, they evolve to several distinct values.}
\label{fig:ring}
\end{figure*}
We consider a two-tunneling model of $N$ site chain, with a single pair of gain and loss potentials $\pm i\gamma$ at locations $m_0$ and $\bar{m}_0$. The tunneling amplitude for sites between the gain and the loss is given by $J$, whereas the tunneling amplitude elsewhere, including between the first and the $N$th site is given by $J'$. The $\mathcal{PT}$ threshold for such a two-tunneling chain is given by $\gamma_{PT}=|J-J'|$~\cite{derekring}. In particular, when $J'=0$, we get an open chain with gain and loss potentials at the two ends, and when $J'=J$, due to the doubly degenerate spectrum of the resulting ring, the $\mathcal{PT}$ threshold is suppressed to zero. 

Figure~\ref{fig:ring} shows the fate of adjacent-site phase differences $\theta_k(t)$. Figure~\ref{fig:ring}a shows that deep in the $\mathcal{PT}$ broken state, phase differences $\theta_k(t)$ reach multiple, different steady state values. These values are independent of the random initial state, and the gain-loss strength $\gamma$, but are dependent on the lattice size $N=5$ and the gain site location $m_0=1$. 
Panel (b) shows that when the gain location is changed to $m_0=2$, the steady-state values also change. 
Figure~\ref{fig:ring}c-d show the results for an $N=11$ site lattice with gain potential at $m_0=1$ and loss at $\bar{m}_0=N$, and $\gamma=1.5J\sim1.5\gamma_{PT}$. Note that when $J'=0$, we get a uniform open chain and the phase differences $\theta_k(t)$ ($2\leq k\leq 11$) lock at $\pi/2$.  Panel c shows that when a small $J'=0.01J$ is introduced, the saturation values split from $\pi/2$. Panel d shows that for $J'\rightarrow J$, the phase differences $\theta_k(t)$ in the $\mathcal{PT}$ broken state saturate to many different values. 

The results in Fig.~\ref{fig:ring} show that the phase-locking phenomenon remains valid for $\mathcal{PT}$-symmetric rings, although the resultant saturation values are not confined to $\pm\pi/2$. 


\section{Discussion}
\label{sec:disc}
Motivated by the time-invariant $\mathcal{PT}$ product, in this paper, we have investigated the fate of phase-differences between wave function amplitudes on adjacent sites in $\mathcal{PT}$-symmetric lattice models. For a wide variety of open chains, deep in the $\mathcal{PT}$-broken region, we numerically found that $\theta_k(t)\rightarrow3\pi/2$ for all $k\leq m_0$ and $\theta_k(t)\rightarrow\pi/2$ for $k>m_0$, where $m_0$ is the site with the largest gain potential. We also found that this pattern disappears for lattices with periodic boundary conditions, although the phenomenon of saturation remains true. 

Can these results be obtained analytically? Let us consider the $\mathcal{PT}$-symmetric trimer from Sec.~\ref{sec:twothree}. Deep in the $\mathcal{PT}$-broken region, $\gamma/J\gg 1$, the trimer has one amplifying mode $|v_+\rangle$ with eigenvalue $i\Gamma=+i\sqrt{\gamma^2-J^2}$, one neutral mode $|v_0\rangle$ with zero eigenvalue, and one decaying mode $|v_{-}\rangle$ with eigenvalue $-i\Gamma$. In the $\mathcal{PT}$-symmetry broken phase, the $\mathcal{PT}$ operator connects the amplifying and the decaying modes, i.e. $\mathcal{PT}|v_{\pm}\rangle=|v_{\mp}\rangle$ and leaves the modes with real eigenvalues unchanged, $\mathcal{PT}|v_0\rangle=|v_0\rangle$. Since the $\mathcal{PT}$ Hamiltonians we consider are symmetric, $H^T=H$, the left-eigenvectors used to form the bi-orthogonal basis are given by transpose of the corresponding right-eigenvectors. As a result $|v_\mu\rangle^T |v_\nu\rangle\propto\delta_{\mu\nu}$. Note that this means the $\mathcal{PT}$ inner-product of a complex-eigenvalued state with itself is zero, whereas the $\mathcal{PT}$ inner-product of a state with purely real eigenvalue is, in general, nonzero. Expanding a random initial state in the right-eigenvector basis gives 
\begin{eqnarray}
\label{eq:1}
|\psi(t)\rangle &=& c_{+}|v_{+}\rangle e^{\Gamma t}+c_0|v_0\rangle+c_{-}|v_{-}\rangle e^{-\Gamma t},\\
\mathcal{PT}|\psi(t)\rangle & = & c^*_{+}|v_{-}\rangle e^{\Gamma t}+c^*_0 |v_0\rangle +c^*_{-}|v_{+}\rangle e^{-\Gamma t}.
\end{eqnarray}
Due to the bi-orthogonality constraints, the $\mathcal{PT}$ inner-product reduces to the following time-invariant expression, $\langle\psi(t)|\psi(t)\rangle_{PT}=|c_0|^2|v_0\rangle^T|v_0\rangle+c_{+}c^{*}_{-}|v_{+}\rangle^T|v_{+}\rangle+c_{-}c^*_{+}|v_{-}\rangle^T|v_{-}\rangle$. This analysis shows that, even in the deep $\mathcal{PT}$-broken region, it is not sufficient to approximate the time-evolved state with its projection onto the fastest-amplifying mode. In addition, Eq.(\ref{eq:npt3}) refers only to difference between the wave-function phases on the first the and third sites, but has no reference to the wave-function phase on the central site. In general, for an $N$-site case, there are $(N-1)$ phase differences $\theta_k\equiv\left[\phi_k-\phi_{k-1}\right]$ . The $\mathcal{PT}$ time invariant provides one equation among them, but additional equations (in the form of other time-invariants) are needed to obtain the phase-locking results that we have numerically obtained. 

This raises the following question: how many linearly independent time-invariants are there? This is an open question. For any Hermitian, intertwining operator $\eta$, i.e. an operator that satisfies $\eta H_{PT}=H^\dagger_{PT}\eta$, it is easy to show that the overlap $\langle\phi(t)|\eta|\psi(t)\rangle$ remains time-invariant; in particular, when $H=H^\dagger$, $\eta=1$ gives the invariance of the Dirac inner product in traditional quantum theory. For an $N$-dimensional Hamiltonian $H_{PT}\neq H_{PT}^\dagger$, a full characterization of the intertwining operators $\eta$ remains an open problem; its solution, most likely, will provide a way to analytically understand the phase-locking patterns that we have numerically discovered. 


\acknowledgments
YJ thanks Frantisek Ruzicka for useful discussions and the anonymous referee for very useful comments. This work was supported by National Science Foundation grant no. DMR 1054020.


\end{document}